\documentclass[letterpaper]{emulateapj}
\slugcomment{\sc Accepted to the Astronomical Journal}
\usepackage{graphicx}

\def\mr{\mathrm}
\def\ergsec{\mr{~erg~s}^{-1} }

\def\kms{\mr{~km\ s}^{-1} }
\def\hi{HI~}
\def\mhi{M_{\hi}}
\def\msol{\mr{M}_{\odot}}
\def\kmsmpc{\mr{\ km}\ \mr{s}^{-1}\ \mr{Mpc}^{-1}}
\def\kmpc{\mr{~Mpc}^{3}}
\def\mpc{\mr{~Mpc}}
\def\jks{\mr{~Jy~km~sec}^{-1}}
\begin{document}
\title{\hi Observations of Five Groups of Galaxies}
\author{E. Freeland\altaffilmark{1,2}, A. Stilp\altaffilmark{3}, E. Wilcots\altaffilmark{1}}

\altaffiltext{1}{Department of Astronomy, University of Wisconsin, 475 N. Charter St., Madison, WI 53706, USA}
\altaffiltext{2}{Email: freeland@astro.wisc.edu}
\altaffiltext{3}{Department of Astronomy, University of Washington, Box 351580, Seattle, WA 98195, USA}


\begin{abstract}
We present the results of \hi observations of five groups of galaxies spanning a range of velocity dispersion and spiral fraction (brightest optical group member in parenthesis): NGC 7582 (NGC 7552), USGC U207 (NGC 2759), USGC U070 (NGC 664), USGC U412 (NGC 3822), USGC U451 (NGC 4065).  Neutral intragroup gas is detected in three of the five groups.  We present the discovery of a previously uncataloged galaxy in the USGC U070 group at $\alpha(2000)= 01^\mr{h} 45^\mr{m} 27^\mr{s}$, $\delta(2000) = +04\ 36\arcmin 19\arcsec$ which we are designating FSW J014526.92+043619.1.  We compile an \hi mass function for the group environment and find that the faint-end slope is consistent with being flat.

\end {abstract}


\section{Introduction}

The majority of galaxies in the local universe reside in groups
\citep{1983ApJS...52...61G, 1987ApJ...321..280T, 2004MNRAS.348..866E,
2007ApJ...671..153Y,2008A&A...479..927T}.   Early group samples were defined as
overdense areas on the sky from photographic plates while redshift surveys
later allowed for groupings in a third dimension.  Investigations have sought
to determine whether these systems are bound gravitationally or simply chance
projections of unrelated galaxies along a line of sight.  Observations of
groups enveloped in a hot, X-ray emitting halo similar to the hot medium found
in rich clusters \citep{2003ApJS..145...39M} along with evidence in others for
interactions between group members have shown that many groups are indeed
gravitationally bound systems.

Galaxy properties, like morphology and star-formation rate (SFR), are observed to change with environment; high density regions, like clusters, are characterized by a larger fraction of elliptical galaxies and a lower star-formation rate than seen in the field \citep{2003MNRAS.346..601G,2003ApJ...584..210G}.  These observations are known as the morphology-density and SFR-density relations, although \citet{1993ApJ...407..489W} argue that these reflect a tighter and more fundamental morphology-radius relation.  These relations are not limited to clusters, they are seen in groups of galaxies as well \citep{2003MNRAS.339L..29H} which suggests that the physical mechanisms which play a role in producing these relations are also important in the group environment.  These mechanisms include some combination of tidal interactions, mergers, ram pressure stripping, and strangulation.  Although, there is debate over the effectiveness of ram pressure stripping and strangulation in the group environment \citep{2004ApJ...616..192C,2008ApJ...672L.103K,2008MNRAS.388.1245R}.  We use \hi because it is an excellent tracer of tidal interactions and often shows evidence for these gravitational disturbances when optical observations of the same source appear undisturbed.  Additionally, \hi observations of groups are able to probe a population of gas rich galaxies which may not be detected in optical observations.  

Our current understanding of the evolution of groups is derived from observed correlations between the fraction of a group's galaxies that are elliptical, the presence of a diffuse X-ray emitting halo, and a high velocity dispersion \citep{1998ApJ...496...39Z,2003ApJS..145...39M}.  Additionally, numerical simulations suggest that galaxies within a compact group eventually merge to form large ellipticals \citep{1997MNRAS.286..825A}. Thus, the evolution of a group should be marked by the removal of \hi from the individual galaxies, dumping it into the intragroup medium where it can be heated, along with primordial intergalactic gas, to X-ray emitting temperatures.  This may be initially accompanied by enhanced star formation within the individual galaxies that leads to the feedback of diffuse hot gas into the intragroup medium \citep{2008MNRAS.391..110D,1994PhDT........29M}.  

We use these data to compile an HI mass function (HIMF) for galaxy groups.  In general, the least well-known component of the HIMF is the slope of the low mass end.  Cosmological and galaxy evolution models depend on this slope as a constraint.  A handful of blind \hi surveys, as well as targeted HI observations of individual clusters and groups of galaxies, have attempted to accurately determine the slope of the low mass end of the \hi mass function \citep{2003AJ....125.2842Z,2002ApJ...567..247R}.  Evidence from these, and other, observations suggests that the low mass slope of the HIMF may vary with environment \citep{2005ApJ...621..215S,2005MNRAS.359L..30Z}.    

In this paper we examine the \hi content of a diverse sample of galaxy groups.  Galaxy morphologies are taken from NED when possible and if not, HyperLeda.  A Hubble constant of $H_0 = 75 \kmsmpc$ is used throughout this paper.
\begin{deluxetable*}{lccccccccc}
\tabletypesize{\scriptsize}
\tablecaption{Group Data\label{tab:groupdata}}
\tablehead{
\colhead{Group}	& \multicolumn{2}{c}{Beam size} &  \colhead{$\sigma(S_\nu)$}  & \colhead{$3\sigma(N_{\mr{\hi}})$}  & \colhead{$3\sigma(\mr{M}_{\odot})$}  & \colhead{$\sigma_{\mr{v}}$} & \colhead{$\log{L_x}$} & \colhead{$f_{sp}$} & \colhead{N}\\
		& ($\arcsec \times \arcsec$)& (kpc $\times$ kpc)& ($\mr{mJy\ beam^{-1}}$)   	 & ($10^{18}\mr{cm}^{-2})$ & ($10^8 \ \mr{M}_{\odot}$)  & ($\kms$) & ($\mr{ergs\ s^{-1}}$) &  & }
\startdata
NGC 7582         &62 $\times$ 49&6.41 $\times$ 5.54& 0.9/3.8      & 13   & 0.12 & $38^{+37}_{-36}$    &$< 40.74$ &1.00 &8 \\ 
USGC U207	&62 $\times$ 60&25.5 $\times$ 24.7& 0.58/0.75	 & 11	& 2.2  & 122 	             &$< 41.90$	&0.86  &7 \\
USGC U070 	&67 $\times$ 60&23.9 $\times$ 21.4& 0.78/0.98    & 6.6  & 0.92 & $130^{+244}_{-99}$  &$< 40.98$ &1.00  &5 \\ 
USGC U412	&67 $\times$ 60&24.0 $\times$ 21.5& 0.26/0.37	 & 4.7	& 0.78 & 168                 &$< 40.88$	&0.90  &10 \\  
USGC U451	&67 $\times$ 56&29.3 $\times$ 24.2& 0.48/0.62	 & 8.7	& 1.6  & 416$\pm$35          &\phn42.40 &0.31  &74   
\enddata
\tablecomments{Sensitivities are calculated using the value for the flux density in the center of the field of view, other relevant assumptions are indicated in \S 3.1.  The final column, N, gives the number of members in the group.  Velocity dispersions for NGC 7582 and USGC U070 are from ZM98, USGC U412 and USGC U207 are from \citet{2002AJ....123.2976R}, and USGC U451 is from \citet{2004ApJ...607..202M}.  X-ray luminosities are from \citet{2003ApJS..145...39M} for NGC 664, NGC 7582, and USGC U412.  The USGC U207 X-ray luminosity is from \citet{2000ApJ...534..114M} and USGC U451 is from \citet{2004MNRAS.350.1511O}.  Spiral fractions are calculated as the ratio of number of spiral galaxies to total number of group members and include S0 galaxies.}
\end{deluxetable*}

\section{The Sample}
 
The five groups in this paper are found in various catalogs including, but not
limited to, Geller and Huchra (GH) \citep{1983ApJS...52...61G}, UZC-SSRS2
(USGC) Group Catalog \citep{2002AJ....123.2976R}, the ROSAT All-Sky Survey
(RASS) Center for Astrophysics (CfA) Loose System (RASSCALS) catalog
\citep{2000ApJ...534..114M}, the Group Evolution Multi-wavelength Study (GEMS)
project sample \citep{2004MNRAS.350.1511O}, and \citet{1998ApJ...496...39Z}, a
spectroscopic survey of poor groups, hereafter known as ZM98.  We identify most
groups in this paper by their USGC number, although alternate identifications
are available in the table notes for each group.  All of these catalogs, except
the GEMS sample and ZM98, use the standard friends-of-friends algorithm
outlined in Huchra and Geller \citep{1982ApJ...257..423H} with parent catalogs
like the Center for Astrophysics (CfA) redshift survey, the Updated Zwicky
Catalog (UZC), and the Southern Sky Redshift Survey (SSRS2) to identify groups.
These parent catalogs have a limiting magnitude of $m_{B(0)}\simeq15.5$.  The
GEMS sample uses the NASA Extragalactic Database (NED) as a parent catalog and
only accepts galaxies as group members if they are within these spatial and
velocity constraints: a projected group radius, $r_{500}$, calculated from the
X-ray temperature of the group and a velocity range of $v\pm3\sigma_{\mr{v}}$.

We chose the five groups discussed in this paper in order to sample group
environments spanning a range of spiral fraction.  However the spiral fraction
as well as the group velocity dispersion can vary, for a given group,
between catalogs and may change as additional redshift information allows for
the identification of new group members.

\section{Observations}
\subsection{\hi Observations}
The data for USGC U412, U451, and U207 were taken with the VLA in D configuration (longest baseline 1030m) in the spring of 2003.  In order to cover the largest velocity range possible we used the correlator in 4 IF mode and centered each IF pair at well separated velocity overlapping the three central channels which gave a $\sim 5.2$ MHz or $1100 \kms$ bandwidth with a velocity resolution of $97.7$ kHz or $21.5 \kms$.  During processing we combined the IFs into one continuous cube using the AIPS task UVGLU.  We used three by three and two by two pointing mosaics resulting in spatial coverage of $1^\circ$ by $1^\circ$ (or 45$\arcmin$ by 45$\arcmin$) on the sky.  Pointing centers were separated by the VLA half-power beamwidth of 15$\arcmin$ and we integrated for 5 minutes on each pointing, shifting to the phase calibrator after each $\sim 45$ minute mosaic sequence.  

USGC U070 was observed with the VLA in the D configuration during August 2004.  The correlator was configured in the standard 2 IF mode with a bandwidth of 3.1 MHz or $660 \kms$ and channel size of 48.4 kHz or $10.3 \kms$.  A three by three pointing mosaic was used to cover the size of the group on the sky.  We used AIPS to calibrate all the VLA data. 

The NGC 7582 group was observed using the Australia Telescope Compact Array (ATCA) near Narrabri in New South Wales, Australia.  We used a 23 pointing mosaic to cover an area of $1.5^\circ \times  1.5^\circ$ on the sky.  We obtained 60 hours of data in the 0.750A configuration (longest baseline 3750m) in January 2002 and 50 hours of data in the EW352 configuration (longest baseline 4439m) in June 2002.  For these observations the ATCA correlator collected dual-polarization data with a total bandwidth of 16 MHz or $3378 \kms$ and a channel width of 31.2 kHz or $13.2 \kms$.  The primary calibrator used to set the absolute flux scale was the standard ATCA flux calibrator 1934-638 (14.9 Jy at 1384 MHz; Reynolds 1994) and was observed for approximately 10 minutes every 24 hours.  A secondary calibrator 2323-407 was observed for about 5 minutes every hour for phase calibration.  MIRIAD was used for editing, calibration, and imaging of the visibility data.  

With calibrated data in hand we used MIRIAD to mosaic, subtract the continuum, and deconvolve the dirty cubes for all the groups.  The naturally weighted cubes were deeply cleaned using MOSSDI, the cleaning task for mosaics.  Line-free channels were identified, and MIRIAD task UVLIN was used to subtract the continuum in the uv plane.  Moment maps were made by smoothing the cube by twice the spatial resolution and masking that cube at the three sigma noise level.  This smoothed, masked cube was then used as a mask when taking the moment of the original data cube.

Additionally, calibrated ATCA data on the NGC 7582 group were obtained in collaboration with B. Koribalski at Australia Telescope National Facility.  These data are single pointings in the 375 (longest baseline 5969m), 750D (longest baseline 4469m), and 1.5A (longest baseline 4469m) configurations, adding a total of 100 more observing hours on this group.  

Table \ref{tab:groupdata} illustrates the basic properties of the five groups and their data cubes.  Mosaicing produces an area of uniform sensitivity where the primary beams overlap surrounded by an area with increased noise.  Column 4 gives the rms noise (in mJy $\mr{beam^{-1}}$) for the center and edge of the field in a line free channel of the deconvolved, naturally weighted cube.  Column density and mass sensitivity calculations correspond to a three sigma detection at the lower rms noise level from the center of the field.  The column density sensitivity is from one line free channel while the mass sensitivity assumes a detection over three channels filling one beam. 
\begin{deluxetable*}{lcccccc}[ht]
\tabletypesize{\scriptsize}
\tablecaption{NGC 7582 Group\tablenotemark{a} : \hi Detections\label{tab:n7582}}
\tablehead{
\colhead{Galaxy}  & \colhead{$\alpha$}  & \colhead{$\delta$} & \colhead{NED} & \colhead{NED velocity} & \colhead{$S_{\mr{\hi}}$} & \colhead{$\mr{M_{\hi}}$}\\
  & (J2000)& (J2000) &Morphology & $(\kms$) & ($\jks$) & ($10^9\ \mr{M}_{\odot}$)}
\startdata
ESO 347-G 008   	& 23 20 49 & -41 43 51  & SAB(s)m 	& 1622 $\pm$ 76 & 8.6 & 0.94 \\
NGC 7599  		& 23 19 21 & -42 15 25	& SB(s)c      	& 1654 $\pm$ \phn7  & 35 & 3.8 \\
NGC 7590 		& 23 18 55 & -42 14 21  & S(r?)bc Sy2 	& 1596 $\pm$ \phn7  & 20 & 2.1 \\
NGC 7582  		& 23 18 24 & -42 22 14  &(R'\_1)SB(s)ab Sy2& 1575 $\pm$ \phn7  & 21 &2.3 \\
LCRS B231454.2-423320 	& 23 17 39 & -42 16 55  & -	        & 1609 $\pm$ 80	    & 0.33 & 0.036 \\
NGC 7552  		& 23 16 11 & -42 34 59  & (R')SB(s)ab;HIILINER & 1585 $\pm$ \phn5   & 36 & 3.9
\enddata
\tablenotetext{a}{Also known as the Grus Group.}
\end{deluxetable*}

\subsection{Optical Data}
Optical data were obtained for USGC U412, U451, and U207 with the WIYN 0.9m telescope on Kitt Peak in Arizona during March 2003 to search for optical counterparts to the \hi detections.  Groups were imaged in B and R filters using the CCD Mosaic imager.  The field of view of this instrument on the WIYN 0.9m telescope covers $1^\circ$ by $1^\circ$ and we used two pointing offsets to fill in the gaps between chips.  These images were reduced using the \emph{mscred} package in IRAF.  Standard calibration procedures were used including overscan and cross talk corrections, bias subtraction, and flat fielding.  Once we had calibrated single images, three per pointing, we stacked them to remove the chip gaps and produce a final image.
For USGC U070 and the NGC 7582 group we use Digitized Sky Survey (DSS) optical data to search for counterparts.

\section{Discussion of Individual Groups}

\subsection{NGC 7582}
The NGC 7582 group, also known as the Grus Group, has eight spectroscopically confirmed members in ZM98 all of which are classified as late-type galaxies.  In ZM98 the morphological classification of group members was done by examination of images and assignments were simply given as early (E and S0) or late (all others).  The group velocity is $1612\pm46 \kms$ and the velocity dispersion is $38\pm37 \kms$ (ZM98).  {\it ROSAT} failed to detect the presence of any diffuse X-ray emission \citep{2003ApJS..145...39M} as did XMM-Newton \citep{2006ApJ...646..143F} with an exposure of 17.8 ksec giving an upper limit of $\log{L_x}<41.05 \ergsec$.  

\begin{figure}
\plotone{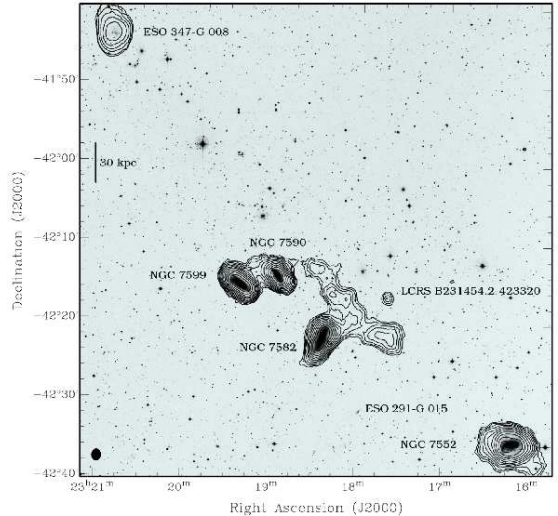}
\caption{Total intensity \hi contours overlaid on a red DSS image of the NGC 7582 group.  The lowest contour corresponds to a three sigma detection at a column density of $3.5 \times 10^{19}\ \mr{cm^{-2}}$ for all sources except ESO 347-G 008 whose lowest contour is $2.2 \times 10^{20}\ \mr{cm^{-2}}$, contours increase by $\sqrt{2}$.  The beam size is $90\arcsec \times 72\arcsec$ and is shown in the lower left.  All group members are labelled except for NGC 7632 which is beyond the field of view of this image.}
\label{fig:7582}
\end{figure}

We detect \hi in six group galaxies: NGC 7599, NGC 7590, NGC 7582, NGC 7552, ESO 347-G008, and LCRS B231454.2-423320.  We do not detect \hi in the optically compact galaxies ESO 291-G015, which has NED morphology of Sa and is well within the spatial and velocity bounds of our data, and NGC 7632, which has a NED morphology of SO and is near the edge of our field of view.  NGC 7582 and NGC 7552 are classified as active (Seyfert 2) galaxies \citep{1994ApJ...437L..17F,1997ApJ...488..174S,2000ApJ...531..245T} and \hi is seen in absorption against their nuclei \citep{1996ASPC..106..238K}.  Table \ref{tab:n7582} lists the \hi detections and Figure \ref{fig:7582} shows the \hi contours on an optical DSS image of the group.

The compact core of the group shows multiple signatures of interactions including a large tidal extension with an integrated flux density of $10.1 \jks$ or $1.3\times10^9~\msol$ which reaches from NGC 7582 towards NGG 7590 as well as southeast towards NGC 7552.  There may also be an \hi bridge between NGC 7599 and NGC $7590$ but we are not able to resolve this structure.

Our NGC 7582 cube covers a velocity range from $98-3480 \kms$ which corresponds to distances of $1.3-46 \mpc$.  The region of uniform sensitivity on the sky is square with $70 \arcmin$ on a side which corresponds to $0.03 \mpc$ at a $1.3\mpc$ distance, 0.5 Mpc at the distance of the group, and $0.94 \mpc$ at a $46 \mpc$ distance.  Thus, the volume surveyed for this group is $14 \kmpc$. 

\subsection{USGC U207}
\begin{figure}
\plotone{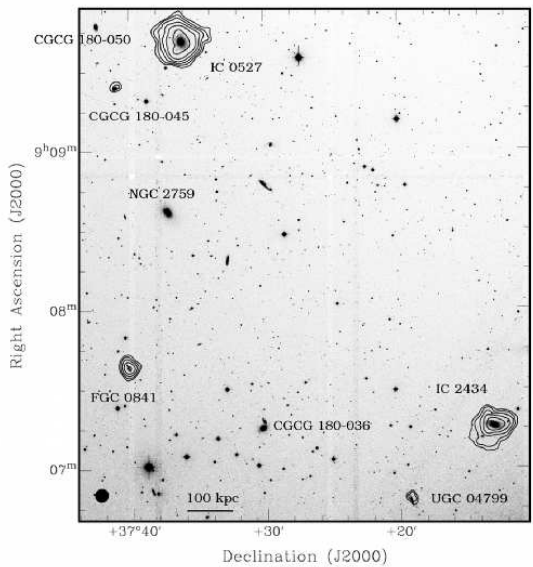}
\caption{Total intensity \hi contours for USGC U207 overlaid on an R band optical WIYN 0.9m image.  The contours correspond to a three sigma detection with the lowest at a column density of $8.2 \times 10^{19}\ \mr{cm^{-2}}$ and increasing by $\sqrt{2}$.  The beam size is $61\arcsec \times 60\arcsec$ and is shown in the lower left.  All group members are labelled except for CGCG 180-048 which is beyond the field of view of this image and our \hi data.}
\label{fig:gh40}
\end{figure}

This group has seven members: IC 0527, IC 2434, UGC 04799, NGC 2759, CGCG 180-036 NED01, CGCG 180-048, and CGCG 180-050.  We detect \hi in IC 0527, IC 2434, UGC 04799 as well as two new galaxies previously not classified as group members: FGC 0841 and CGCG 180-045.  All detections are spiral galaxies except CGCG 180-045 whose morphology is unknown.  CGCG 180-036 is a galaxy pair with discrepant velocities such that CGCG 180-036 NED2 is a background object and not a group member.  CGCG 180-048 is way beyond the spatial bounds of our data.  IC 2434 has an interesting HI morphology that is reminiscent of ram pressure stripping with the classical head-tail morphology showing compressed contours opposite extended contours.  However, in this environment it is much more likely that this \hi morphology is caused by a tidal interaction with nearby galaxy UGC 04799.  The heliocentric group velocity is $7182\pm64 \kms$ and the group velocity dispersion is $\sigma_{\mr{v}}=177 \kms$ \citep{2002AJ....123.2976R}.  \citet{2000ApJ...534..114M} give an upper limit on the X-ray emission from this group of $\log{L_x} < 41.90 \ergsec$.  The dynamic range of our \hi data for this system is limited by a bright continuum source in the field which doesn't subtract cleanly from both IFs.  Table \ref{tab:gh40} lists the \hi detections in this group and Figure \ref{fig:gh40} shows the total intensity HI contours on an optical image of the group.

Our USGC U207 cube covers a velocity range from $6350-7450 \kms$ which corresponds to distances of $85-99 \mpc$.  The region of uniform sensitivity on the sky is square with $40 \arcmin$ on a side which corresponds to $0.99 \mpc$ at a $85 \mpc$ distance, $1.1 \mpc$ at the distance of the group, and $1.2 \mpc$ at a $99 \mpc$ distance.  Thus, the volume surveyed for this group is $17 \kmpc$.

\begin{deluxetable*}{lccccccccc}
\tabletypesize{\scriptsize}
\tablecaption{USGC U207\tablenotemark{a} : \hi Detections\label{tab:gh40}}
\tablehead{ 
\colhead{Galaxy} & \colhead{$\alpha$}  & \colhead{$\delta$} & \colhead{NED} &  \colhead{NED velocity} & \colhead{$S_{\mr{\hi}}$} & \colhead{$\mr{M_{\hi}}$}   \\
		 & (J2000)              & (J2000)             & Morphology    & ($\kms$) & ($\jks$) & ($10^9\ \mr{M}_{\odot}$) }
\startdata 
IC 0527        	& 9 09 41.8   & +37 36 05  & S? & 6889 $\pm$ 11 & 9.8  & 20   \\ 
CGCG 180-045   	& 9 09 24.3   & +37 41 09  & -  & 7138 $\pm$ 65 & 0.49 & 0.97  \\
UGC 04799      	& 9 08 47.6   & +37 30 06  & SB & 6873 $\pm$ 27 & 0.55 & 1.1  \\	
FGC 0841       	& 9 07 38.5   & +37 40 21  & Sc & 6976 $\pm$ 24 & 1.9  & 3.8   \\
IC 2434        	& 9 07 16.0   & +37 12 55  & SB?& 7158 $\pm$ 29 & 4.6  & 9.4   
\enddata
\tablenotetext{a}{Also known as RASSCALS NRGb019 (7 members).}
\end{deluxetable*}

\pagebreak
\subsection{USGC U070}
\begin{figure}
\plotone{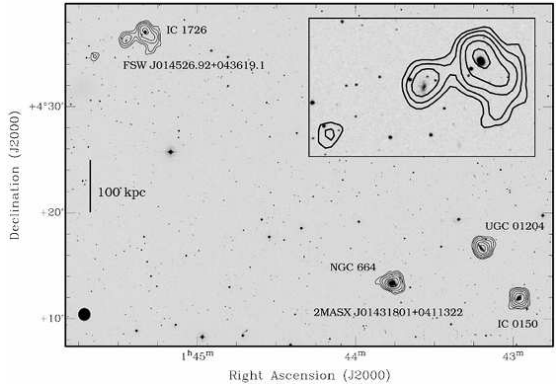}
\caption{Total intensity \hi contours for the USGC U070 group overlaid on a blue DSS image.  Newly detected galaxy, FSW J014526.92+043619.1, is shown, along with IC 1726, in an enlarged panel in the upper right.  Near FSW J014526.92+043619.1 we detect $4.4\times10^8~\msol$ of \hi without an optical counterpart.  The contours correspond to a five sigma detection with the lowest at $8.7\times 10^{19}\ \mr{cm^{-2}}$  and increasing by $\sqrt{2}$.  The beam size is $67\arcsec \times 60\arcsec$ and shown in the lower left corner.  All group members are labelled except for NGC 664:[ZM98] 057 (NED identification) which is beyond the field of view of this image and our \hi data.}
\label{fig:664}
\end{figure}

USGC U070, also know as the NGC 664 group, has five spectroscopically identified members in ZM98: NGC 664, IC 0150, and UCG 01204 are classified as spirals while 2MASX J01431801+0411322, and NGC 664:[ZM98] 057 (NED identification) have unknown morphologies.  The group velocity is $5399\pm80 \kms$ and the group velocity dispersion is reported to be $130^{+244}_{-99}$ (ZM98).  \citet{2003ApJS..145...39M} give an upper limit of $40.98 \ergsec$ for diffuse X-ray emission from the intragroup medium using {\it ROSAT} PSPC data with an exposure time of 7.5 ks.  We detect \hi in the three group members previously classifed as spirals, as well as two new galaxies: IC1726 and its uncataloged neighbor ($\alpha(2000)= 01^\mr{h} 45^\mr{m} 27^\mr{s}$, $\delta(2000) = +04\ 36\arcmin 19\arcsec$) which we are designating FSW J014526.92+043619.1 according to IAU Nomenclature conventions.  Additionally, we detect $4.4\times10^8~\msol$ of \hi near FSW J014526.92+043619.1 without an optical counterpart which may be tidal material or an \hi cloud.  We do not detect \hi in either of the two group members with unknown morphologies.  Although NGC 664:[ZM98] 057 and 2MASX J01431801+0411322 both have emission lines in their spectra \citep{1998ApJ...496...39Z}, NGC 664:[ZM98] 057 is spatially beyond the coverage of our \hi data and 2MASX J01431801+0411322 is optically very compact compared to the other galaxies in this group.  Its \hi mass may fall below our detection limit.  Table \ref{tab:664} lists the \hi detections in this group and Figure \ref{fig:664} shows the total intensity \hi contours on an optical image of the group with an inset showing the uncataloged galaxy. 

Our NGC 664 cube covers a velocity range from $5120-5750 \kms$ which corresponds to distances of $68-77 \mpc$.  The region of uniform sensitivity on the sky is square with $40 \arcmin$ on a side which corresponds to $0.79 \mpc$ at a $68 \mpc$ distance, $0.85 \mpc$ at the distance of the group, and $0.89 \mpc$ at a $77 \mpc$ distance.  Thus, the volume surveyed for this group is $6.0 \kmpc$.

\begin{deluxetable*}{lcccccc}
\tabletypesize{\scriptsize}
\tablecaption{USGC U070\tablenotemark{a} : \hi Detections\label{tab:664}}
\tablehead{
\colhead{Galaxy} & \colhead{$\alpha$}  & \colhead{$\delta$} & \colhead{NED} &\colhead{$v$} & \colhead{$S_{\mr{\hi}}$} & \colhead{$\mr{M_{\hi}}$} \\
		 & (J2000)             & (J2000)            & Morphology    & ($\kms$) & ($\jks$)       & ($10^9\ \mr{M}_{\odot}$)}
\startdata
IC 0150			& 01 42 58 & +04 11 58 & Sb  & 5551 $\pm$ 38 &  2.6 & 3.4 \\
UGC 01204		& 01 43 12 & +04 16 42 & Sbc & 5621 $\pm$ 41 &  2.8 & 3.6 \\
NGC 664		 	& 01 43 46 & +04 13 22 & Sb  & 5424 $\pm$ \phn5  &  2.9 & 3.7 \\
IC 1726			& 01 45 20 & +04 37 06 & -   & 5334 $\pm$ 36 &  3.1 & 4.0 \\
FSW J014526.92+043619.1 & 01 45 27 & +04 36 16 & -   & 5320 $\pm$ 10\tablenotemark{b}  &1.0 & 1.3\\ 
\hi cloud               & 01 45 39 & +04 34 43 & -   & 5230 $\pm$ 20\tablenotemark{b}   & 0.34& 0.44
\enddata
\tablenotetext{a}{Also known in ZM98 as the NGC 664 Group (6 members).}
\tablenotetext{b}{Velocity from \hi data.}
\end{deluxetable*}

\subsection{USGC U412}
\begin{figure}
\plotone{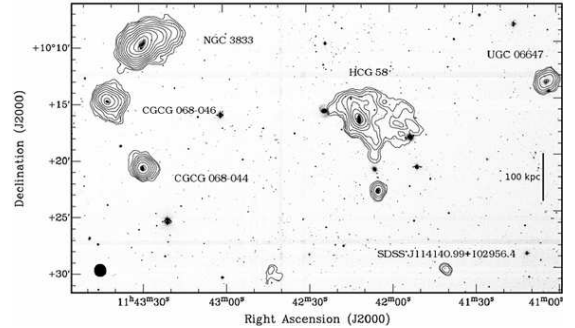}
\caption{Total intensity \hi contours for the USGC U412 group overlaid on an R band optical WIYN 0.9m image.  The contours correspond to a three sigma detection with the lowest at a column density of $3.5 \times 10^{19}\ \mr{cm^{-2}}$ and increasing by $\sqrt{2}$.  The beam size is $66\arcsec \times 60\arcsec$ and is shown in the lower left.  All group members are labelled except for NGC 3839 and IC 0727 which are beyond the field of view of this data.}
\label{fig:gh89}
\end{figure}

This group has ten spectroscopically identified members in the UZC-SSRS2 Group Catalog; nine of them are classified as spiral galaxies.  The group velocity is $6154 \kms$ and the velocity dispersion is $168 \kms$ \citep{2002AJ....123.2976R}.  We detect \hi in five of these galaxies, namely NGC 3833, CGCG 068-046, CGCG 068-044, NGC 3822, UGC 06647.  We also detect two new group members in \hi: NGC 3820, which was not originally classified as a member of USGC U412 although it is a member of the compact subgroup HCG 58, and SDSS J114140.99+102956.4.   CGCG 068-044, CGCG 068-046, and NGC 3833 have large \hi disks extending from three to five times the optical radius of the galaxy.  NGC 3819, the only early type group member, is not detected, while NGC 3839 and IC 0727 are undetected because they are outside the field of view of our \hi data.  The apparent lack of an \hi detection for NGC 3825 will be discussed in the next section with the \hi content of HCG 58.  Table \ref{tab:gh89} lists the \hi detections in this group and Figure \ref{fig:gh89} shows the total intensity \hi contours on an optical image of the group.  

Our USGC U412 cube covers a velocity range from $5550-6650 \kms$ which corresponds to distances of $74-89 \mpc$.  The region of uniform sensitivity on the sky is square with $25 \arcmin$ on a side which corresponds to $0.54 \mpc$ at a $74 \mpc$ distance, $0.6 \mpc$ at the distance of the group, and $0.65 \mpc$ at a $89 \mpc$ distance.  Thus, the volume surveyed for this group is $5.2 \kmpc$.

\begin{deluxetable*}{lcccccc}
\tabletypesize{\scriptsize}
\tablecaption{USGC U412\tablenotemark{a} : \hi Detections\label{tab:gh89}}
\tablehead{
\colhead{Galaxy} & \colhead{$\alpha$}  & \colhead{$\delta$} & \colhead{NED} &  \colhead{NED velocity} & \colhead{$S_{\mr{\hi}}$}    & \colhead{$\mr{M_{\hi}}$}   \\
  & (J2000)& (J2000)& Morphology  &   ($\kms$) & ($\jks$)              & ($10^9\ \mr{M}_{\odot}$)  }
\startdata 
CGCG 068-046   	&11 43 41.7  & +10 14 44  & Sc    &      6009 $\pm$ 18 & 3.4  &5.3 \\
CGCG 068-044   	&11 43 29.3  & +10 20 42  & Sc    &      6013 $\pm$ 19 & 2.1  &3.2 \\
NGC 3833       	&11 43 28.9  & +10 09 43  & Sc    &      6036 $\pm$ 10 & 13  &19 \\
NGC 3822 + extended&11 42 11.1  & +10 16 40  & Sb    &   6138 $\pm$ 20 & 9.0  & 15 \\
NGC 3820       &11 42 04.9  & +10 23 03  & Sbc   &       6102 $\pm$ 33 & 0.68 & 1.0 \\
SDSS J114140.99+102956.4 & 11 41 41.0 & +10 29 56 &  - & 6163 $\pm$ 48 & 0.31  &0.48 \\
UGC 06647 	&11 41 03.7 	& +10 13 31 & Sd  &      6350 $\pm$ 66   & 1.7  &2.6 
\enddata
\tablenotetext{a}{Also known as GH 89 (7 members), RASSCALS NRGb151 (12 members), GEMS HCG058 (7 members), and MKW10 (9 members).}
\end{deluxetable*}

\subsubsection{HCG 58}
Most Hickson Compact Groups (HCGs) are dynamically associated with larger loose groups \citep{2001RMxAA..37..173T}.  This may explain the observed longevity of these systems when compared with N-body simulations which show an isolated compact group evolving into a single merger remnant in only a few orbital times \citep{1989Natur.338..123B}.  HCG 58 \citep{1982ApJ...255..382H} appears to be at the core of the previously discussed loose group, USGC U412.  Hickson finds five members in HCG 58, four of which are classified as spirals.  We detect \hi in three of them, NGC 3820, NGC 3817, and NGC 3822.  There is also an \hi extension towards NGC 3825 but very little \hi coincident with the optical galaxy.  The last group member, NGC 3819, classified as E1 is undetected.  HCG 58 is classified as \hi deficient by \citet{2001A&A...377..812V} who find a total \hi mass of $6.8 \times 10^9 \msol$ compared to our total mass of $1.6\times 10^{10} \msol$ which leaves little room for deficiency.   

\begin{figure}
\plotone{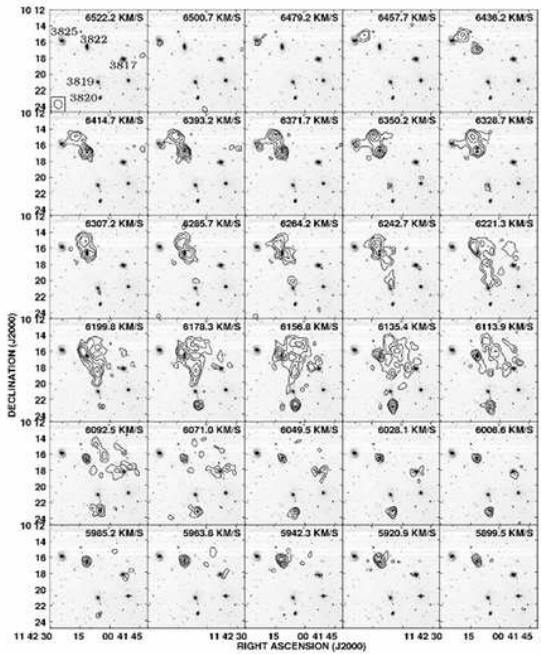}
\caption{Channel map with \hi contours for HCG 58 overlaid on an R band optical WIYN 0.9m image.  The contours correspond to a three sigma detection with the lowest at a column density of $6.5 \times 10^{18} \mr{cm^{-2}}$ and increasing by $\sqrt{2}$.  The beam size is $66\arcsec \times 60\arcsec$ and is shown in the lower left.  Group members are labelled in the first panel by their NGC designation.}
\label{fig:hcg58}
\end{figure}

Figure \ref{fig:hcg58} shows the \hi emission (contours on an optical image of HCG 58) from this group as a function of velocity.  The \hi morphology of NGC 3822 is severely disturbed.  We see an \hi extension towards NGC 3825 which contains a similar amount of \hi as seen in previous single dish measurements ($0.80 \pm 0.15 \jks$) \citep{1989gcho.book.....H}.  It appears that NGC 3825 has had most of its \hi stripped off into the intragroup medium due to recent interactions with the other group members.  The published X-ray data for this group do not agree on whether a hot intragroup medium exists.  \citet{1994AJ....107..427D} find an X-ray luminosity, whose profile is dominated by a point source, of $L_x = 0.90 \times 10^{42}\ \mr{ergs\ sec}^{-1}$ using the Einstein X-ray satellite, while the RASSCALS study \citep{2000ApJ...534..114M} finds $L_x = 1.349 \times 10^{42}\ \mr{ergs\ sec}^{-1}$ using ROSAT All-Sky Survey data.  However, using only pointed-mode ROSAT data \citet{2003ApJS..145...39M} do not detect X-rays in HCG 58 and put upper limits on the X-ray luminosity at $L_x < 7.586 \times 10^{40}\ \mr{ergs\ sec}^{-1}$.  Nonetheless, \citet{2008ApJ...685..858F} have shown that the intragroup medium in even small groups can be relatively dense and remain undetected by the current generation of X-ray telescopes.

\subsection{USGC U451}

This group has the highest velocity dispersion of the five groups presented in
this paper and is the only group with a definitive detection of X-ray emitting
intragroup gas.  A redshift survey by \citet{2004ApJ...607..202M} finds 74
group members, a group velocity of $6995\pm48\kms$, and a group velocity
dispersion of $416 \kms$.   The group has two compact subgroups, UZC-CG 156
(including NGC 4076) and UZC-CG 157 (including NGC 4098) which were identified
by the 3D UZC catalog solely on the criterion of compactness
\citep{2002A&A...391...35F} unlike HCGs which were visually selected using
multiplicity, isolation, and luminosity concordance of member galaxies. 

\citep{2000MNRAS.315..356H} detect bimodal X-ray emission with one peak on the galaxies NGC 4061 and NGC 4065 and the other on NGC 4066.  The bimodal X-ray emission does not correlate with the two compact subgroups.  The GEMS \citep{2004MNRAS.350.1511O} and RASSCALS \citep{2000ApJ...534..114M} X-ray luminosities are in close agreement with $\log L_x=42.64\pm0.05$ and $\log L_x=42.54\pm0.08$ respectively.  

\begin{figure}
\plotone{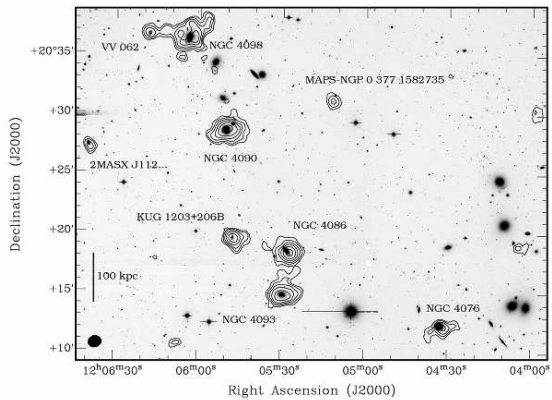}
\caption{Total intensity \hi contours for the USGC U451 group overlaid on an R band optical image taken at the WIYN 0.9m.  The center of the group, including the X-ray emission, is near the two elliptical galaxies (NGC 4065 and NGC 4061) in lower right corner of the image.  The contours correspond to a three sigma detection with the lowest at a column density of $5.9\times10^{19} \mr{cm^{-2}}$ and increasing by $\sqrt{2}$.  The beam size is $68\arcsec \times  56\arcsec$ and is shown in the lower left.  The center of the group is located at 12h04m09.5s +20d13m18s (J2000).  We have only spatially sampled one quarter of the group.  This group has many members so we have labeled only the \hi detected galaxies.}
\label{fig:gh98}
\end{figure}

We detect \hi in VV 062, NGC 4092, NGC 4086, 2MASX J12063969+2027235, and MAPS-NGP O 377 1582735 and place lower limits on the \hi content of NGC 4098, NGC 4090, NGC 4076, and KUG 1203+206B.  NGC 4098 is a system of two interacting spiral galaxies which is itself interacting with the pair of spiral galaxies, VV062a and VV062b.  NGC 4086 and NGC 4092 are connected by an \hi bridge.  At the time of the observations the VLA did not have the bandwidth necessary to sample the entire velocity range of the group which spans from $\sim 5800-7800 \kms$ \citep{2004ApJ...607..202M}.  Additionally, our nine pointing VLA mosaic has only spatially sampled one quarter of the group size.  Table \ref{tab:gh98} lists the \hi detections in this group and Figure \ref{fig:gh98} shows the total intensity \hi contours on an optical image of the group.  Galaxies whose \hi profiles are truncated by the edge of our data cube are indicated with lower limits in Table \ref{tab:gh98}. 

Our USGC U451 cube covers a velocity range from $6000-7200 \kms$ which corresponds to distances of $80-96 \mpc$.  The region of uniform sensitivity on the sky is square with $25 \arcmin$ on a side which corresponds to $0.58 \mpc$ at a $80 \mpc$ distance, $0.67 \mpc$ at the distance of the group, and $0.70 \mpc$ at a $96 \mpc$ distance.  Thus, the volume surveyed for this group is $6.6 \kmpc$.

\begin{deluxetable*}{lcccccc}                                                     
\tabletypesize{\scriptsize}                                                     
\tablecaption{USGC U451\tablenotemark{a} : \hi Detections\label{tab:gh98}}                           
\tablehead{                                                                     
\colhead{Galaxy} & \colhead{$\alpha$}  & \colhead{$\delta$} & \colhead{NED} &   \colhead{NED velocity}& \colhead{$S_{\mr{\hi}}$}    & \colhead{$\mr{M_{\hi}}$}  \\            
  & (J2000)& (J2000)& Morphology  &  ($\kms$) & ($\jks$)  & ($10^9\ \mr{M}_{\odot}$) }                     
\startdata							      
2MASX J12063969+2027235 &  12 06 39.7  & +20 27 24  & S? & 7263 $\pm$ 16 	&\phn0.79 & 1.8\\
VV 062                  &  12 06 18.3  & +20 36 40  & Sb  &   7217 $\pm$ 18 	&\phn0.85 & 1.8  \\
NGC 4098                &  12 06 03.9  & +20 36 22  & Sbc  & 7296 $\pm$ 11 	&$>4.4$ & $>9.9$  \\
NGC 4092		 &  12 05 50.1  & +20 28 37  & Sa  & 6719 $\pm$ \phn8 	&\phn1.6 & 3.0 \\
KUG 1203+206B		 &  12 05 47.4  & +20 19 35  & S  & 6984 $\pm$ 15	 & $>1.1$ & $>2.2$ \\
NGC 4090		 &  12 05 27.9  & +20 18 32  & Sab &  7333 $\pm$ 45	&$>2.8$ & $>6.4$  \\
NGC 4086 		&  12 05 29.4  & +20 14 48  & S0  &  7093 $\pm$ 24	 &\phn2.2 & 4.7  \\
MAPS-NGP O\_377\_1582735  &  12 05 11.8  & +20 31 07  & - &7100\tablenotemark{b}  &\phn2.3 &4.9  \\
NGC 4076 		&  12 04 32.5  & +20 12 18  & Sa  & 6214 $\pm$ 6 	&$>1.6$  & $>2.5$ 
\enddata   
\tablenotetext{a}{Also known as RASSCALS NRGb 177 (16 members), GEMS NGC 4065 (9 members), GH 98 (7 members).}
\tablenotetext{b}{Velocity from \hi data.}
\end{deluxetable*}

\section{Discussion}
\subsection{Intragroup \hi}
The NGC 7582, NGC 664, and USGC U412 groups are the most spiral rich groups and they all contain varying amounts of intragroup \hi.  This material is tidally stripped \hi from the galaxies themselves.  These tidal interactions, which are so favorable in the group environment, are likely to play an important role in the evolution of galaxies in these systems.  As is seen here, they have the potential to remove large amounts of neutral gas from the galaxies involved.  This tidally stripped \hi may fall back onto the galaxies or eventually be photoionized by the UV background \citep{2000AJ....119.1130H}. 

The two systems with the most intragroup \hi, NGC 7582 and HCG 58 (a subgroup of USGC U412), are both compact cores of larger systems.  These compact configurations with small impact parameters and small relative velocities between galaxies are where tidal forces will be most efficient at removing \hi from inside galaxies into the intragroup medium.  \citet{2001A&A...377..812V} examine the \hi content of compact groups of galaxies and find that $70\%$ of spiral galaxies in compact systems have a perturbed \hi morphology.  They present an evolutionary scenario that begins with the least evolved groups showing little evidence for interactions among group members.  Intermediately evolved groups have tidal \hi features and evolved groups may have a single \hi cloud enveloping the group or show extreme deficiencies in galaxy \hi content.  \citet{2008AIPC.1035..201B} find evidence for a significant reservoir of diffuse \hi in compact systems that is likely a result of the tidally stripped \hi being dispersed throughout the intragroup medium.  They find a larger component of this diffuse \hi in compact groups with \hi deficient galaxies.  Within groups of galaxies, the formation of compact subgroups may facilitate interactions and lead to the tidal removal of significant amounts of \hi.  The loss of neutral gas from these galaxies may then restrict their star-forming potential.

\subsection{\hi Mass Function}

A handful of recent large \hi surveys have attempted to accurately determine the slope of the low mass end of the \hi mass function (HIMF).  Predictions from cold dark matter models of galaxy formation suggest that the faint end slope should be rather steep \citep{1999ApJ...524L..19M}.  If these dark matter haloes contain \hi then the corresponding HIMF should also have a steep faint end slope.  Cosmological and galaxy evolution models use the behaviour of the \hi mass function at the low mass end as a constraint.  
We use the data in this paper, as well as additional \hi data in our possession for the following groups: HCG 22 \citep{2003AAS...203.5001K}, NGC5846 and NGC 2563 \citep{wilcots09prep} to compile an \hi mass function for the group environment.  These additional data were taken as mosaics with the VLA and reduced in a similar manner to the data presented in this paper.  We do not have enough data to fit an independent mass function but we do check the consistency of our mass function against previous HIMF results.

\begin{figure}
\plotone{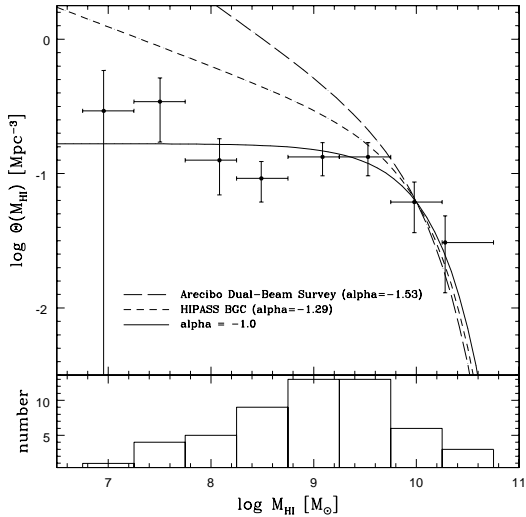}
\caption{The combined \hi mass function for nine galaxy groups overlaid with Schechter functions fit from larger surveys as well as a flat function.  Upper Panel: Horizontal error bars indicate bin widths, vertical error bars indicate Poisson noise.  Each point is plotted according to the average of the \hi masses within that bin.  The HIPASS Bright Galaxy Catalog (BGC) \hi mass function is from Zwaan et al. 2003 and the Arecibo Dual-Beam Survey \hi mass function is from \citep{2002ApJ...567..247R}.  A flat HIMF with faint end slope $\alpha=-1.0$ is plotted for reference.  The data are not consistent with either the HIPASS or ADB survey mass functions.  Lower Panel:  The number of detections per mass bin. }
\label{fig:himf}
\end{figure}

The HIMF, shown in Figure \ref{fig:himf}, is the space density of \hi objects in units of $h_{75}^3\ \mr{Mpc}^{-3}$.  It is parametrized by a Schechter function of the form
\begin{equation} \theta\left(M\right) =\theta^* \ln\left(10\right) \left(\frac{\mhi}{M_*}\right)^{\alpha+1} \exp\left(-\frac{\mhi}{M_*}\right) \end{equation}
where $M_*$ is the mass at which the function turns over, $\alpha$ is the slope of the low mass end, and $\theta^*$ is a normalization.

Recent large \hi surveys have allowed for the determination of the HIMF with increased precision.  \citet{2005MNRAS.359L..30Z} use 4315 galaxies from the \hi Parkes All-Sky Survey (HIPASS) to construct an HIMF and derive a low mass slope of $\alpha=-1.37$.  With 265 galaxies the Arecibo Dual-Beam Survey (ADBS) \citep{2002ApJ...567..247R} finds an even steeper faint end slope of $\alpha=-1.53$.  \citet{2002ApJS..142..161P} carried out a sensitive search for \hi companions to otherwise isolated spiral galaxies.  Their survey covered a total of 31.6 Mpc$^{3}$ around 41 different spirals, and the faint end slope of their HIMF is consistent with values flatter than $\alpha\le-1.3$. 

There is some evidence for variation in the HIMF with environment.  Using only galaxies near the Virgo cluster \citet{2002ApJ...567..247R} find a faint end slope that is much more flat ($\alpha=-1.2$) than what is seen in the full sample of galaxies.  \citet{2002A&A...382...43D}  showed that the \hi mass functions in both the Centaurus A and Sculptor groups are inconsistent with a slope steeper than $\alpha=-1.5$.  \citet{2005nfcd.conf..351K} perform a deep \hi survey of the Canis Venatici group and report a flat ($\alpha = -1.07$) HIMF.  Another flat HIMF is seen in a deep survey of the Ursa Major group by \citet{2000ASPC..218..263V}. \citet{2005ApJ...621..215S} use 2771 \hi detections from an optically selected sample of galaxies to derive an HIMF whose faint-end slope is $\alpha=-1.24$ for the full sample and see evidence for a flattening of the HIMF when examined in increasingly dense environments.  However, \citet{2005MNRAS.359L..30Z} using HIPASS data find tentative evidence for a steepening of the faint-end slope of the HIMF in increasingly rich environments.  They parametrize the density around a given galaxy by calculating the distance to its n nearest HIPASS detected neighbors.  The flattening of the faint end slope of the HIMF suggests that physical processes in the group environment are responsible for removing gas from low mass galaxies.

Large surveys must deal with volume corrections which account for the fact that low luminosity objects are more easily detected nearby and their contribution throughout the whole volume must be weighted accordingly.  These corrections must also account for the local overdensity that surrounds the Milky Way.  Since all galaxies in a group are at approximately the same distance no correction needs to be made for large scale structure effects throughout the depth of the survey.  We do apply a volume correction, discussed below, to account for low mass galaxies which are not detectable throughout an entire data cube.  \citet{2004ApJ...607L.115M} explore the impact of distance uncertainties on the faint-end slope of mass and luminosity functions.  They use mock \hi survey data to show that for galaxies in the Virgo region the low-mass HIMF slope is underestimated if peculiar velocities are neglected when finding the distances to objects in the nearby universe.  Objects with $cz > 6000 \kms$ and typical peculiar velocities of a few hundred kilometers per second are well within a pure Hubble flow and have errors of less than $10\%$ on their redshift distances.  We have not made any correction for the groups in our sample with $cz < 6000 \kms$ because they are all in the anti-Virgo direction where the local peculiar velocity field is not well constrained.

Low \hi mass galaxies which can be detected near the front of the cube may be undetectable at the back side of the cube.  This means that we are sensitive to low \hi masses in a subset of the total volume surveyed.  To account for this we apply a volume correction to the lowest three mass bins \citep{1968ApJ...151..393S}.  To determine this correction for a given cube we calculate the actual volume over which we could detect a galaxy with a given mass based on the noise level and size of the cube.  Then we add the volumes calculated by this method together for all cubes.  The HIMF is then produced by dividing the total number of detections in a mass bin by the corresponding volume surveyed for masses in that bin among all groups.   

Our \hi mass function for the group environment, seen in Figure \ref{fig:himf}, is essentially flat and inconsistent with the HIPASS and ADBS \hi mass functions.  Our HIMF is similar to those compiled for other group environments.  The volume corrections for the last two bins are very large with only $3\%$-$10\%$ of the total volume being surveyed for galaxies with those masses.  Of the ten galaxies in the last three bins six of them (including all the galaxies in the last two bins) are detections from galaxy group NGC 5846 \citep{wilcots09prep}.  However, the flattening of the low mass slope is apparent even neglecting the last three bins.  A large and sensitive survey that can populate the lower mass bins and whose detections can be reliably sorted by environment will be necessary to address the variety currently seen in the HIMF. 

\section{Summary}

We have performed mosaicked VLA and ATCA \hi observations of five groups of galaxies.  We find a multitude of interactions and interesting \hi morphologies in these systems.  Intragroup \hi, not easily associated with a single group member, is present in the three most spiral rich systems.  An \hi mass function compiled from these data, and an additional three groups, is consistent with other \hi mass functions for the group environment in that it has a flat low mass slope \citep{2000ASPC..218..263V,2002A&A...382...43D}.  The flattening of the faint end slope of the HIMF suggests that physical processes in the group environment are responsible for removing gas from low mass galaxies.

\begin{acknowledgements} We would like to thank the referee for useful comments and gratefully acknowledge the support of NSF grant AST-0506628.  EF acknowledges the support of the Wisconsin Space Grant Consortium in the form of a Graduate Fellowship.  This research has made use of the NASA/IPAC extragalactic database (NED) which is operated by the Jet Propulsion Laboratory, Caltech, under contract with the National Aeronautics and Space Administration.  The National Radio Astronomy Observatory is a facility of the National Science Foundation operated under cooperative agreement by Associated Universities, Inc. 
\end{acknowledgements}


\end{document}